%% file: BABAR-PUB-08057.tex
\documentclass[twocolumn,showpacs,aps,prd,floatfix]{revtex4}
\usepackage{graphicx}
\usepackage{dcolumn}
\usepackage{epsfig}
\usepackage{psfrag}
\usepackage{amsmath}
\usepackage{longtable}
\usepackage{bm}
\usepackage{relsize}
\usepackage{xspace}

\long\def\inst#1{\par\nobreak\kern 4pt\nobreak
    {\it #1}\par\vskip 10pt plus 3pt minus 3pt}
\def\babar{\mbox{\slshape B\kern-0.1em{\smaller A}\kern-0.1em
    B\kern-0.1em{\smaller A\kern-0.2em R}}}
\def\pep2{PEP-II}
\def\Kp    {\ensuremath{K^+}\xspace}
\def\Km    {\ensuremath{K^-}\xspace}
\def\piz   {\ensuremath{\pi^0}\xspace}
\def\pip   {\ensuremath{\pi^+}\xspace}
\def\pim   {\ensuremath{\pi^-}\xspace}
\def\Dz{D^0}
\def\KS    {\ensuremath{K^0_{\scriptscriptstyle S}}\xspace}
\def\D       {\ensuremath{D}\xspace}
\def\Dp      {\ensuremath{D^+}\xspace}
\def\Dm      {\ensuremath{D^-}\xspace}
\def\Dz      {\ensuremath{D^0}\xspace}
\def\Dzb     {\ensuremath{\Dbar^0}\xspace}
\def\DzDzb   {\ensuremath{\Dz {\kern -0.16em \Dzb}}\xspace}
\def\DpDm    {\ensuremath{\Dp {\kern -0.16em \Dm}}\xspace}
\def\Dstar   {\ensuremath{D^*}\xspace}
\def\Dstarb  {\ensuremath{\Dbar^*}\xspace}
\def\Dstarz  {\ensuremath{D^{*0}}\xspace}

\def\Dstarzb {\ensuremath{\Dbar^{*0}}\xspace}
\def\Dstarp  {\ensuremath{D^{*+}}\xspace}
\def\Dstarm  {\ensuremath{D^{*-}}\xspace}

\def\psiprpr  {\ensuremath{\psi(3770)}\xspace}
\def\BR         {{\ensuremath{\cal B}\xspace}}
\def\invfb   {\ensuremath{\mbox{\,fb}^{-1}}\xspace}
\def\Dbar    {\kern 0.2em\overline{\kern -0.2em D}{}\xspace}
\def\Db      {\ensuremath{\Dbar}\xspace}
\def\DDb     {\ensuremath{\D {\kern -0.16em \Db}}\xspace}
\def\DDbX     {\ensuremath{\D {\kern -0.16em \Db}X}\xspace}
\def\DDstarb     {\ensuremath{\Dstar}\ensuremath{\Dbar}\xspace}
\def\DstarDstarb     {\ensuremath{\Dstar{\kern -0.16em \Dstarb}}\xspace}
\def\DstarzDstarzb {\ensuremath{D^{*0}}\ensuremath{\Dbar^{*0}}\xspace}
\def\DstarpDstarm  {\ensuremath{D^{*+}}\ensuremath{D^{*-}}\xspace}
\def\DstarzDzb {\ensuremath{D^{*0}}\ensuremath{\Dbar^0}\xspace}
\def\DstarpDm  {\ensuremath{D^{*+}}\ensuremath{D^{-}}\xspace}
\def\Dstarzzb {\ensuremath{\Dbar^{(*)}}\xspace}
\def\Y#1S{\ensuremath{\Upsilon{(#1S)}}\xspace}
\def\FourS {\Y4S}
\def\MM{M^2_{\rm rec}}
\def\MDstar{M_{\Dstar}}
\def\MD{M_{\D}}

\newcommand{\gevc}{\ensuremath{{\mathrm{\,Ge\kern -0.1em V\!/}c}}\xspace}
\newcommand{\mevc}{\ensuremath{{\mathrm{\,Me\kern -0.1em V\!/}c}}\xspace}
\newcommand{\gevcc}{\ensuremath{{\mathrm{\,Ge\kern -0.1em V\!/}c^2}}\xspace}
\newcommand{\mevcc}{\ensuremath{{\mathrm{\,Me\kern -0.1em V\!/}c^2}}\xspace}
\newcommand{\mev}{\ensuremath{\mathrm{\,Me\kern -0.1em V\!}}\xspace}
\newcommand{\gev}{\ensuremath{\mathrm{\,Ge\kern -0.1em V\!}}\xspace}
\newcommand{\gevcccc}{\ensuremath{{\mathrm{\,Ge\kern -0.1emV^2\!/}c^4}}\xspace}

\begin{document}
\newcommand{\BaBarYear}    {08}
\newcommand{\BaBarNumber}  {057}
\newcommand{\BABARProcNumber} {\phantom{14}}
\newcommand{\SLACPubNumber} {13560}
\newcommand{\LANLNumber} {yymm.nnnn [hep-ex]}
\newcommand{\BaBarType}      {PUB}  
\begin{flushleft}
\babar-\BaBarType-\BaBarYear/\BaBarNumber \\
SLAC-PUB-\SLACPubNumber 
\end{flushleft}

\title{\boldmath Exclusive Initial-State-Radiation Production of the \DDb, \DDstarb, and \DstarDstarb Systems}

\input{authors_dec2008.tex}

\begin{abstract}

We perform a study of the 
exclusive production of \DDb, \DDstarb, and \DstarDstarb in initial-state-radiation events, from
$e^+ e^-$ annihilations at a center-of-mass energy near 10.58 \gev, to search for 
charmonium and possible new resonances.
The data sample corresponds to an integrated luminosity of 384~\invfb
and was recorded by the \babar\ experiment at the \pep2 storage rings.
The \DDb, \DDstarb, and \DstarDstarb mass spectra show clear evidence of several $\psi$ resonances.
However, there is no evidence for $Y(4260)\to \DDstarb$ or $Y(4260)\to \DstarDstarb$. 
\end{abstract}


\maketitle
\section{Introduction}
The surprising discovery of new states decaying to  $J/\psi \pi^+ \pi^-$
\cite{charmonium,babar_Y} has renewed interest in the field of charmonium spectroscopy, since the new
resonances are not easy to accommodate in the quark
model.
In particular, the \babar\ experiment discovered a new broad state, $Y(4260)$,
decaying to $J/\psi \pi^+ \pi^-$ in the initial-state-radiation (ISR)
reaction $e^+ e^- \to \gamma_{ISR} Y(4260)$. 
The quantum numbers  $J^{PC}=1^{--}$
are inferred from the single virtual-photon production mechanism.
Further structures at $4.36\ \gevcc$~\cite{babar_Y2,belle_Y2} and $4.66\ \gevcc$~\cite{belle_Y2}
have been observed in the $\psi(2S) \pi^+ \pi^-$ mass distribution from the reaction
$e^+ e^- \to \gamma_{ISR} \psi(2S) \pi^+ \pi^-$. 
Charmonium
states at these masses would be expected \cite{eichten,barnes} to decay predominantly to \DDb, \DDstarb, or
\DstarDstarb~\cite{conj}. 
It is peculiar that the decay rate to the hidden charm final state
$J/\psi \pi^+ \pi^-$ is much larger for the $Y(4260)$ than for excited
charmonium states~\cite{mo},
and that at the $Y(4260)$ mass the cross section for $e^+e^-\to\text{hadrons}$ 
exhibits a local minimum~\cite{pdg}.
Several theoretical interpretations for the $Y(4260)$ have been proposed, 
including unconventional scenarios:  
quark-antiquark gluon hybrids~\cite{Y4260-hybrid}, baryonium~\cite{qiao}, tetraquarks~\cite{maiani},
and hadronic molecules~\cite{Y4260-molecule}. 
For a discussion and a list of 
references see, for example, Ref.~\cite{swanson}.

This work explores ISR production of $\DDb$, $\DDstarb$, and $\DstarDstarb$ final states for evidence of 
charmonium states and 
unconventional structures. This follows an earlier \babar \  measurement of the \DDb cross section~\cite{babar_dd}.
A study by the Belle collaboration of 
the \DDb, \DDstarb, and
\DstarDstarb final states can be found in Refs.~\cite{belle_dd,dstar2}. Recent measurements of the $e^+ e^-$ cross sections
can be found in Ref.~\cite{cleo}.

We also measure for the first time branching fractions of high mass charmonium states, other than $Y(4260)$, for which little
information exists~\cite{pdg}, and compare our measurements with theoretical expectations~\cite{barnes,swanson,eichten}.

This paper is organized as follows. In Section II we give a short description of the \babar \ experiment and in Section III we 
describe the data selection. Section IV is devoted to the selection
of the \DDstarb final state and in Section V, we present the mass resolution, reconstruction efficiency, and measured cross sections.
In Section VI we describe the \DstarDstarb cross section measurement while in Section VII
we present the \DDb data. The description of the fit of the three channels is described in Section VIII, while Section IX is devoted 
to the measurements of the ratios of branching fractions. Finally, in Section X, we compute the limit on production of 
$Y(4260)$ decaying to $\DDstarb$ and \DstarDstarb, and summarize conclusions in Section XI.
  
\section{The \babar\ experiment}
This analysis is based on
 a 384~$\invfb$ data sample recorded at the
\FourS resonance and 40 MeV below the resonance by the \babar\ detector at the \pep2
asymmetric-energy $e^+e^-$ storage rings.  
The \babar\ detector is
described in detail elsewhere~\cite{babar}. We mention here only the parts of the 
detector which are used in the present analysis.
Charged particles are detected
and their momenta measured with a combination of a 
cylindrical drift chamber (DCH)
and a silicon vertex tracker (SVT), both operating within a
1.5 T magnetic field of a superconducting solenoid. 
The information from
a ring-imaging Cherenkov detector combined with energy-loss measurements in the 
SVT and DCH provide identification of charged kaon and pion candidates. 
Photon energies are measured with a 
CsI(Tl) electromagnetic calorimeter.

\section{Data Selection}
$\DDb$ candidates are reconstructed in the seven final states
listed in Table~\ref{tab:list}. 
\begin{table*}[tbp]
\caption{List of the reconstructed $\DDb$ final states.}
\label{tab:list}
\begin{center}
\vskip -0.2cm
\begin{tabular}{llll}
\hline \noalign{\vskip1pt}
N & Channel  & \ \ \ First $D$ decay mode    & \ \ \  Second $D$ decay mode \cr
\hline \noalign{\vskip2pt}
1& $\Dz \Dzb$ & \ \ \ $\Dz \to \Km \pip$ &  \ \ \ $\Dzb \to \Kp \pim$ \cr
2& $\Dz \Dzb$ &\ \ \ $\Dz \to \Km \pip$ &\ \ \ $\Dzb \to \Kp \pim \piz$ \cr
3& $\Dz \Dzb$ & \ \ \ $\Dz \to \Km \pip$ &\ \ \ $\Dzb \to \Kp \pim \pip \pim$ \cr
4& $\Dz \Dzb$ &\ \ \ $\Dz \to \Km \pip \piz$ &\ \ \ $\Dzb \to \Kp \pim \pip \pim$ \cr
5& $\Dp \Dm$ & \ \ \ $\Dp \to \Km \pip \pip$ & \ \ \ $\Dm \to \Kp \pim \pim$ \cr
6& $\Dp \Dm$ &\ \ \ $\Dp \to \Km \pip \pip$ & \ \ \ $\Dm \to  \Kp \Km \pim$ \cr
7& $\Dp \Dm$ &\ \ \ $\Dp \to \Km \pip \pip$ &\ \ \ $\Dm \to \KS \pim$ \cr
\hline
\end{tabular}
\end{center}
\end{table*}
The $\Dstarz \to \Dz \piz$ and $\Dstarz \to \Dz \gamma$
decay modes are used to form $\Dstarz\Dzb$ and $\Dstarz\Dstarzb$ candidates. 
The $\Dstarp \to \Dz \pip$ and $\Dstarp \to \Dp \piz$
decay modes are used to form $\Dstarp\Dm$ and $\Dstarp\Dstarm$ candidates. 
Table~\ref{tab:list_dstar} summarizes
the full decay chains  used to reconstruct the \DDstarb and \DstarDstarb candidates.
\begin{table*}[tbp]
\caption{List of the \DDstarb and $\Dstar\Dstarb$ reconstructed final states. The reconstructed \Dz decay modes are listed in 
Table~\ref{tab:list} for the $\Dstarz\Db$ and $\DstarzDstarzb$ states. The column headed ``Veto'' lists ambiguities 
with the indicated channels, ``Removed'' indicates the fraction of events removed by the veto.}
\label{tab:list_dstar}
\begin{center}
\vskip -0.2cm
\begin{tabular}{rlllcc}
\hline \noalign{\vskip2pt}  
N & Channel   & First decay mode   & \ \  Second decay mode & Veto & \ \ Removed \% \cr
\hline \noalign{\vskip2pt}
8 & \DstarzDzb & $\Dstarz \to \Dz \gamma$ &  & 9-12 & 5.9 \cr
9 & \DstarzDzb & $\Dstarz \to \Dz \piz$ &  & 11,12 & 3.2 \cr
\hline \noalign{\vskip2pt}
10& \DstarzDstarzb & $\Dstarz \to \Dz \gamma$ &\ \  $\Dstarzb \to \Dzb \gamma$ & 9,11 & 1.1 \cr
11& \DstarzDstarzb & $\Dstarz \to \Dz \piz$ &\ \  $\Dstarzb \to \Dzb \gamma$ & 8,10 & 0.7 \cr
12& \DstarzDstarzb & $\Dstarz \to \Dz \piz$ &\ \  $\Dstarzb \to \Dzb \piz$ & &\cr
\hline \noalign{\vskip2pt} 
13& \DstarpDm & $\Dstarp \to \Dz \pip$, $\Dz \to \Km \pip$ &\ \  $\Dm \to \Kp \pim \pim$ & &\cr
14& \DstarpDm & $\Dstarp \to \Dp \piz$, $\Dp \to \Km \pip \pip$ &\ \  $\Dm \to \Kp \pim \pim$ & &\cr
15& \DstarpDm & $\Dstarp \to \Dz \pip$, $\Dz \to \Km \Kp$ &\ \  $\Dm \to \Kp \pim \pim$ & &\cr
16& \DstarpDm & $\Dstarp \to \Dz \pip$, $\Dz \to \Km \pip$ &\ \  $\Dm \to \Kp \Km \pim$ & &\cr
17& \DstarpDm & $\Dstarp \to \Dp \piz$, $\Dp \to \Km \pip \pip$ &\ \  $\Dm \to \Kp \Km \pim$ & &\cr
\hline \noalign{\vskip2pt} 
18& \DstarpDstarm & $\Dstarp \to \Dz \pip$, $\Dz \to \Km \pip$ &\ \  $\Dstarm \to \Dzb \pim$, $\Dzb \to \Kp \pim$  & &\cr
19& \DstarpDstarm & $\Dstarp \to \Dp \piz$, $\Dp \to \Km \pip \pip$ &\ \  $\Dstarm \to \Dzb \pim$, $\Dzb \to \Kp \pim$  & &\cr
\hline
\end{tabular}
\end{center}
\end{table*}

For all final states,
events are retained if the number of well-measured charged tracks, having
a minimum transverse momentum of 0.1 \gevc, is exactly
equal to the total number of charged daughter particles. 
Photons are identified as EMC clusters that do not have a spatial match with a
charged track, and that have a minimum energy of 30 \mev. 
Neutral pion candidates are formed 
from pairs of photons kinematically fitted with the \piz mass constraint. \KS
candidates are
reconstructed, with a vertex fit,  in the $\pip\pim$ decay mode.
The tracks corresponding to the charged daughters of each \D candidate are constrained to come from
a common vertex. Additionally, for the $\Dz \to \Km \pip \piz$ channel, 
the \Dz mass constraint is 
included in the fit, and for the $\Dm \to \KS \pim$ channel, a \KS mass
constraint is imposed. Reconstructed \D candidates with a $\chi^2$ fit 
probability greater than 0.1\% are retained.
Each $\DDb$ pair is refit to a common vertex with
the constraint that the pair originates from the $e^+ e^-$ interaction region.
Only candidates with a $\chi^2$ fit probability greater than 0.1\% are retained.
Background \piz candidates from random combinations of photons and other background channels 
are suppressed by requiring no more
than one \piz candidate other than those attributed to the \Dz and \Dstar decays. Similarly, we 
require in the event no more than one extra photon candidate, having a minimum energy of 100 \mev,
apart from any photon attributed to \Dstar 
or \piz decays. 

For \D decay modes without a \piz daughter, the \D-candidate momentum is determined from the summed 
three-momenta of the decay particles and its energy is computed using the nominal \D mass value~\cite{pdg}.
For the $\Dz \to \Km \pip \piz$ channel, the 4-momentum from the mass-constrained fit is used. 
Similarly, the \Dstar momentum is determined from the summed 
three-momenta of the decay particles and its energy is computed using the nominal \Dstar mass.

The ISR photon is preferentially emitted at small angles 
with respect to the beam axis, and escapes detection in the majority of ISR
events. Consequently, the ISR photon is treated as a missing particle. 
We define the squared mass ($\MM$) recoiling against the \DDb, \DDstarb, and \DstarDstarb systems using the four-momenta of the beam
particles ($p_{e^\pm}$) and of the reconstructed $\D$ ($p_D$) and $\Dstar$ ($p_{\Dstar}$):
\begin{equation}
\MM \equiv ( p_{e^-}+p_{e^+}-p_{D^{(*)}}-p_{\bar D^{(*)}})^2
\end{equation}
This quantity 
should peak near zero for both ISR events and for exclusive production of 
$e^+ e^- \to \DDstarb$ or $e^+ e^- \to \DstarDstarb$. In exclusive production
the \DDstarb and \DstarDstarb mass distributions peak at the kinematic limit. 
Therefore we 
select ISR candidates by requiring \DDb, \DDstarb and \DstarDstarb  invariant masses below 6 \gevcc
and $| \MM |<1 \ \gevcccc$.

\begin{figure}[!htb]
\begin{center}
\includegraphics[width=10cm]{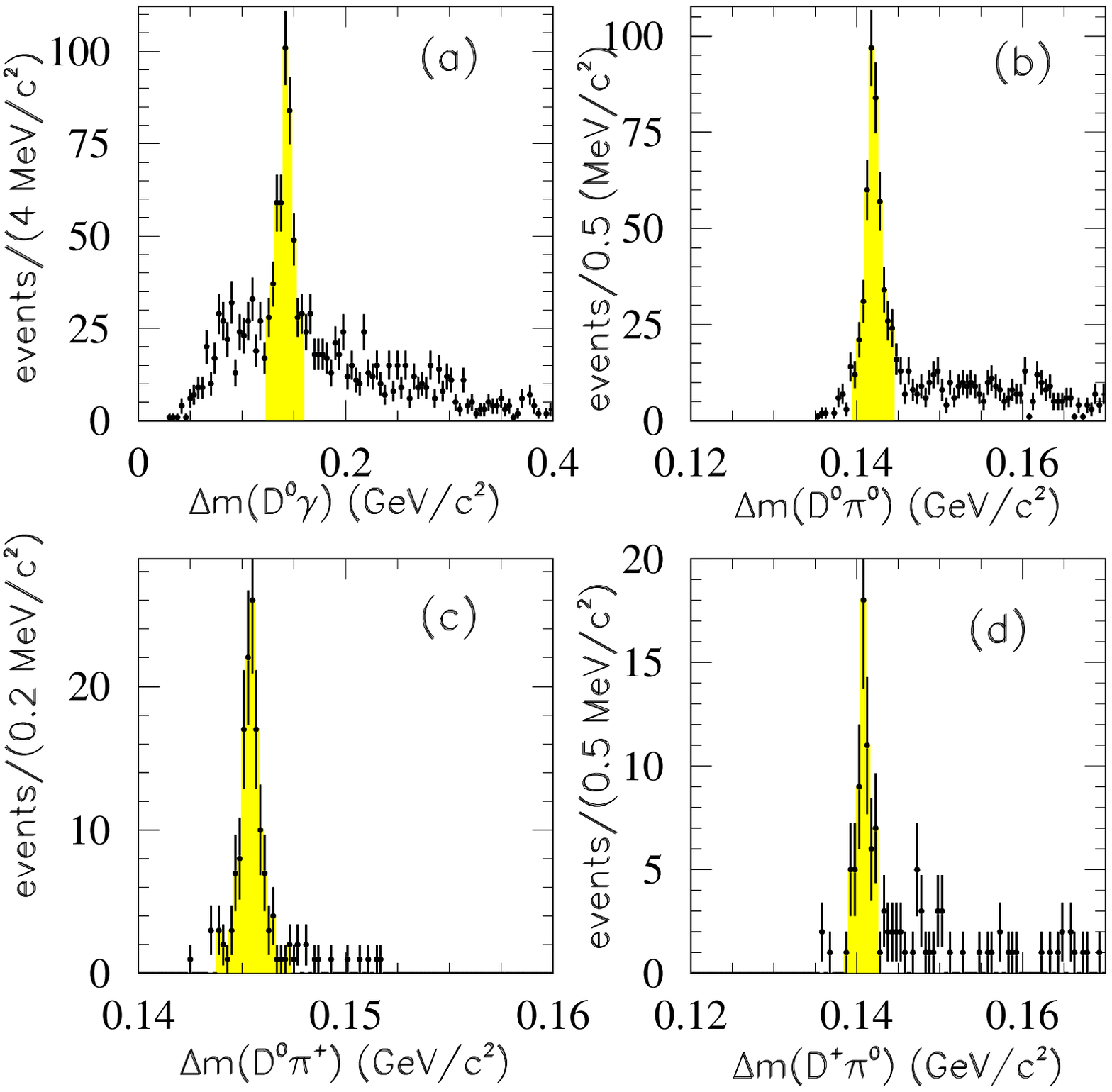}
\caption{$\Delta m$ distributions for \DDstarb candidates after applying the $|\MM |<1\ \gevcccc$ and $m(\DDstarb)<6$ \gevcc selections,
for (a) $\Dstarz \to \Dz \gamma$,
(b) $\Dstarz \to \Dz \piz$, (c) $\Dstarp \to \Dz \pip$ with $\Dz \to \Km \pip$, and
(d) $\Dstarp \to \Dp \piz$ with $\Dp \to \Km \pip \pip$. The shaded regions indicate the ranges used to select
the $D^*$ candidates.}
\label{fig:fig1}
\end{center}
\end{figure}

We select $D$ and $D^*$ candidates based on the candidate $D$ mass and the mass difference 
$\Delta m = \MDstar - \MD$. The  $D$ and $D^*$ 
parameters are obtained by fitting the relevant mass spectra (see Fig.~\ref{fig:fig1} for some
$\Delta m$ distributions) using a polynomial 
for the background and a single Gaussian for the signal. Events are selected within $\pm 2.5 \sigma$ from the 
fitted central values, where $\sigma$ is the Gaussian width. For $\Dstarp \to \Dz \pip$, the selection criterion 
has been extended to $\pm6\sigma$ due to the presence of non-Gaussian tails.

Because of our tolerance of an extra $\piz$ and/or $\gamma$,
an ambiguity can occur for channels involving a $\Dstarz$ which is handled as follows. 
Each combination is considered as a possible candidate for channels 8-12 and \DzDzb.
Monte Carlo simulations weighted by the \DDb, \DDstarb, and \DstarDstarb measured cross
sections~\cite{babar_dd,belle_dd,dstar2} and branching fractions are used to optimize the selection 
criteria and estimate the feedthrough of one channel to the other.
A candidate is rejected if (a) it satisfies all the selection criteria for an  
ambiguous channel and (b) this rejection does not produce any significant loss in the channel
under study and therefore can be classified as background. 
The list of channels rejected in case of ambiguities are indicated in the ``Veto'' column 
in Table~\ref{tab:list_dstar}. The table also lists the fraction of events removed by these cuts in the 
$| \MM |<1 \ \gevcccc$ region. 

In the case of multiple \Dstarz candidates, such as \DstarzDstarzb with both $\Dstarz \to \Dz \gamma$, the candidate with $m(\Dz \gamma)$ 
closest to the nominal \Dstarz mass is accepted.
The charged \DDstarb and \DstarDstarb modes, also listed in Table~\ref{tab:list_dstar},
do not require such a procedure because backgrounds are negligible.

\section{Study of the {\boldmath \DDstarb} final state}

\begin{figure*}[!htb]
\begin{center}
\includegraphics[width=14cm]{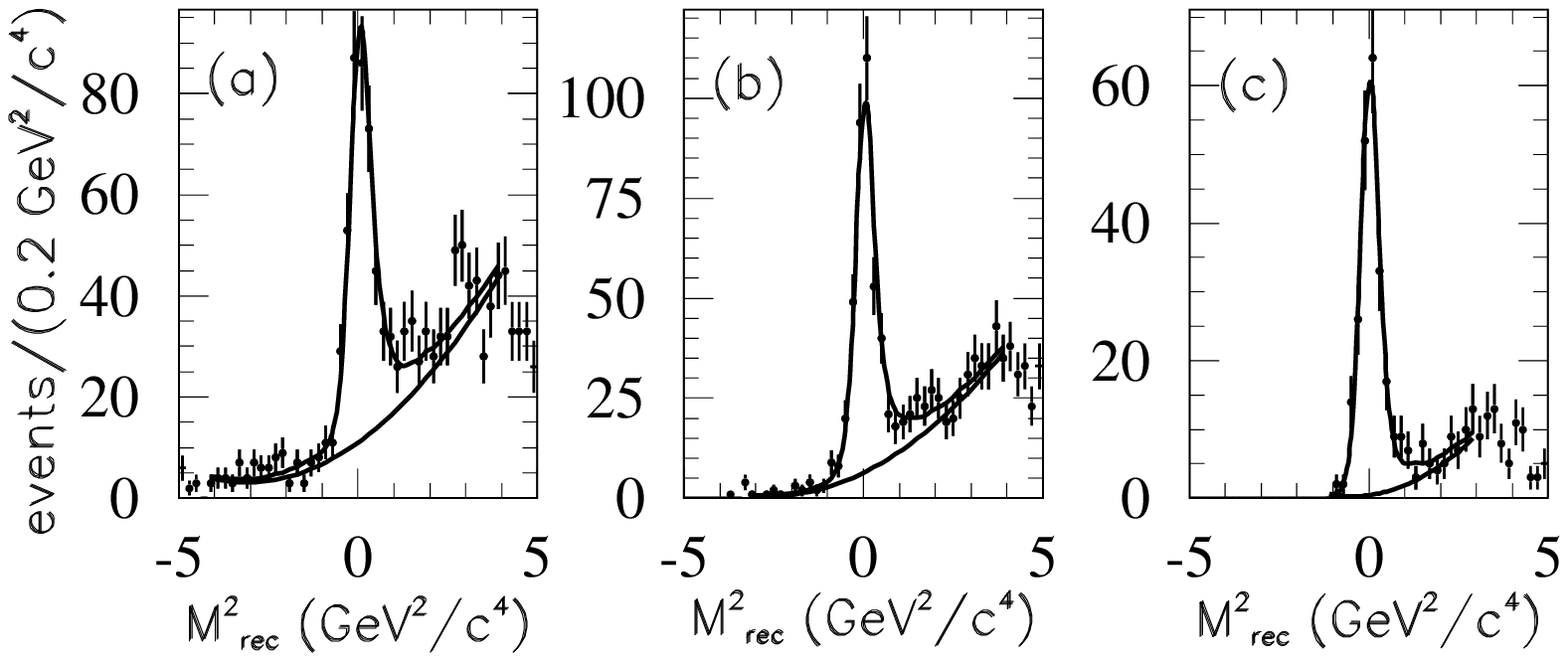}
\caption{Distribution of $\MM$, the mass recoiling against the $\DDstarb$ system, for (a) $\Dstarz \to \Dz \gamma$,
(b) $\Dstarz \to \Dz \piz$, and (c) $\DstarpDm$ candidates. The curves are the results from the fits described in the text.}
\label{fig:fig2}
\end{center}
\end{figure*}

Figure~\ref{fig:fig2} shows the \DDstarb $\MM$ distributions after all the cuts for (a) $\DstarzDzb,\Dstarz \to \Dz \gamma$,
(b) $\DstarzDzb,\Dstarz \to \Dz \piz$, and (c) $\DstarpDm$. Clear peaks centered at zero with little background are observed, providing
evidence of an ISR process.
The number of background events in the $| \MM| <1\ \gevcccc$ is estimated by fitting the $\MM$ distribution for each channel. The fits are performed
using a $2^{nd}$ order polynomial for the background and a signal $\MM$ lineshape obtained from Monte Carlo simulations
corresponding to the relative composition of the data.
The event yields are obtained by subtracting the fitted backgrounds and integrating the resulting $\MM$ distributions in the $| \MM| <1\ \gevcccc$ region. The resulting yields
and fitted purities, defined as $P=N_{\rm signal}/(N_{\rm signal}+N_{\rm background})$, for each channel are summarized in Table~\ref{tab:pury}.

\begin{table}[tbp]
\caption{Number of ISR candidates and purities for the different channels calculated in the range $| \MM| <1\ \gevcccc$.
The last column gives the value of the average efficiency $\epsilon^{\cal{B}}$ at a mass of 4.5 \gevcc.}
\label{tab:pury}
\begin{center}
\vskip -0.2cm
\begin{tabular}{lccc}
\hline \noalign{\vskip2pt}
Channel & Signal+ & Purity(\%) & \ \ $\epsilon^{\cal{B}}\times 10^{-5}$ \cr
 & Background & \cr
\hline \noalign{\vskip2pt}
$\Dz \Dzb$ & 654 & 74.3 $\pm$ 1.7 & \cr
$\Dp \Dm$  & 199 & 88.4 $\pm$ 2.3 & \cr 
\hline \noalign{\vskip2pt}
Total \DDb & 853 & 77.6 $\pm$ 1.4 & 25 \cr
\hline
\hline \noalign{\vskip2pt}
\DstarzDzb, $\Dstarz \to \Dz \gamma$ & 460 & 75.4 $\pm$ 2.0 &\cr
\DstarzDzb $\Dstarz \to \Dz \pi^0$ & 422 & 84.4 $\pm$ 1.8 &\cr
\hline \noalign{\vskip2pt}
Total  \DstarzDzb & 882 & 79.7 $\pm$ 1.4 & 4 \cr
\hline \noalign{\vskip2pt}
\DstarpDm   & 228 & $100_{-3}^{+0}$ & 5 \cr
\hline \noalign{\vskip2pt}
Total  \DDstarb & 1110 & 83.6 $\pm$ 1.1 &  \cr
\hline
\hline \noalign{\vskip2pt}
 \DstarzDstarzb & 293 &  69.3 $\pm$ 3.7 & \cr
\hline \noalign{\vskip2pt}
\DstarpDstarm  & 33 & $100_{-3}^{+0}$ & \cr
\hline \noalign{\vskip2pt}
Total \DstarDstarb  & 326 & 72.1 $\pm$ 2.5 & 1 \cr
\hline
\end{tabular}
\end{center}
\end{table}

The purity of each reconstructed $\Dstar$ channel is also demonstrated in 
Fig.~\ref{fig:fig1}, where
the $\Delta m$ 
distribution is shown for $\DDstarb$ candidates 
with $|  \MM |  < 1\ \gevcccc$ and $\DDstarb$ masses below 6 \gevcc. The final selection
of the ISR candidates is performed applying the $\Delta m$ selection criteria described above.

\begin{figure}[!htb]
\begin{center}
\includegraphics[width=8cm]{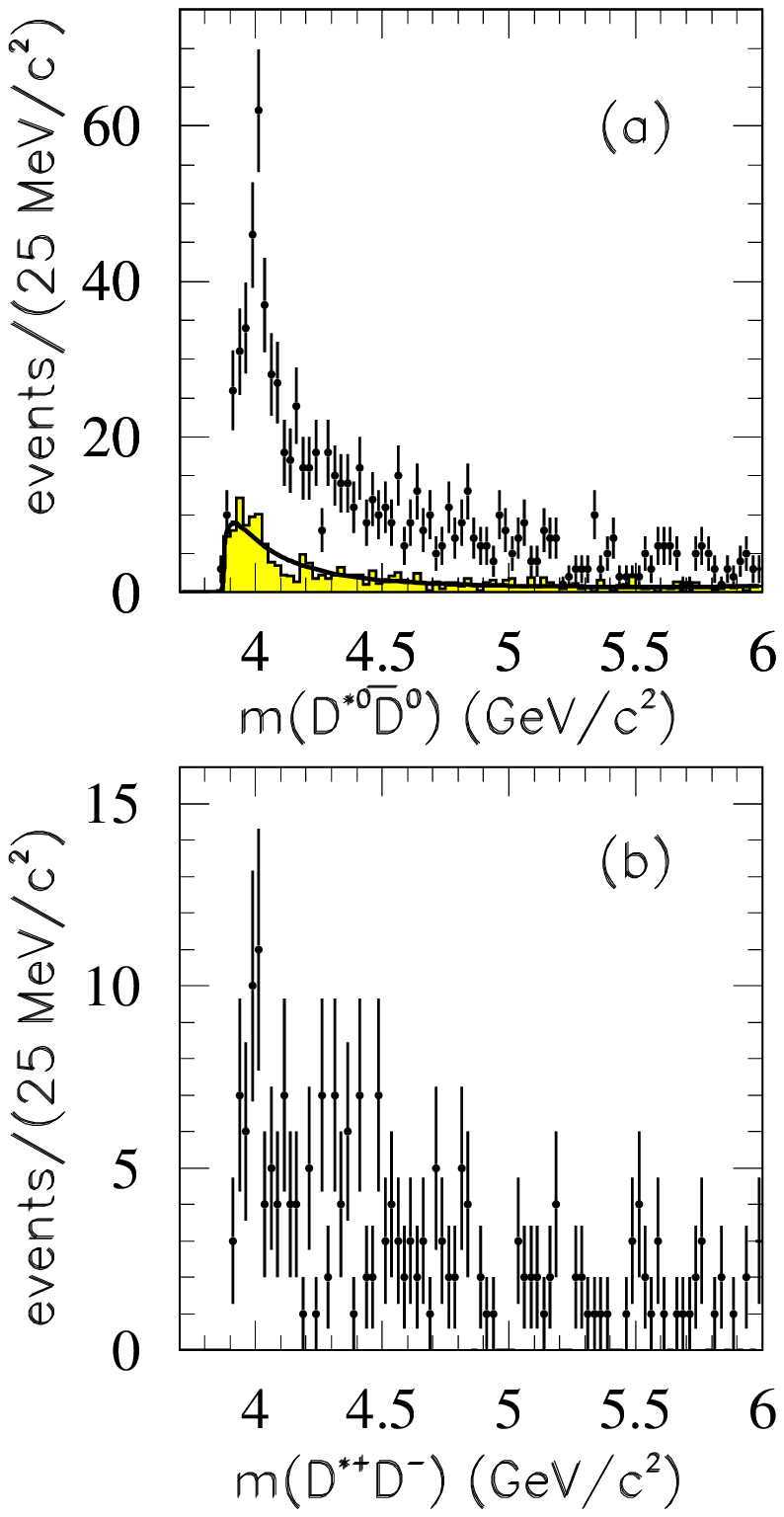}
\caption{(a) \DstarzDzb and (b) \DstarpDm mass spectra. The shaded histogram in (a) is obtained
from the $\MM$ sideband region $1.5 < \MM < 3.5\ \gevcccc$ normalized to the background estimated 
from the fit to the $\MM$ distribution.
The curve is the result from the fit described in the text.} 
\label{fig:fig3}
\end{center}
\end{figure}

The \DstarzDzb mass spectrum is shown in Fig.~\ref{fig:fig3}(a), and the 
\DstarpDm mass spectrum is shown in Fig.~\ref{fig:fig3}(b). Both spectra show an enhancement near threshold
due to the presence of the $\psi(4040)$ resonance.  

The background shape for \DstarzDzb candidates is explored using the $\MM$ sideband region,
$1.5 < \MM < 3.5\ \gevcccc$. The \DstarzDzb mass spectrum for these events, normalized to the 
background estimated from the fit to the $\MM$ distribution, is presented as the shaded histogram
in Fig.~\ref{fig:fig3}(a). This background has been fitted with a threshold function:
\begin{equation}
B(m) = (m - m_{th})^{\alpha + \beta m}e^{-\gamma m - \delta m^2 - \epsilon m^3},
\end{equation}
\noindent where $m_{th}$ is the threshold \DstarzDzb mass. 
The \DstarpDm final state is consistent with having zero background.  

\section{Mass resolution, Efficiency, and {\boldmath \DDstarb} cross section}

In order to measure efficiencies and \DDstarb mass resolutions,
ISR events are simulated at five different values of the \DDstarb invariant
masses between 
4.25 and $6.25\ \gevcc$.
These events are simulated using the
GEANT4 detector simulation package~\cite{geant} and are processed through the same
reconstruction and analysis chain as real events. 
The mass resolution is determined from the difference between generated and
reconstructed \DDstarb  masses. 
The \DDstarb mass resolutions
are similar for all channels and increase with \DDstarb mass from 5 to 10 MeV$/c^2$.

The mass-dependent reconstruction efficiency for channel $i$, $\epsilon_i(m_{\DDstarb})$, is parameterized
by a second-order polynomial, and is multiplied by the decay branching fraction $\BR_i$~\cite{pdg} to define 
\begin{equation}
\epsilon_i^{\cal{B}}(m_{\DDstarb}) = \epsilon_i(m_{\DDstarb}) \times \BR_i.
\end{equation}
These values are weighted by $N_i(m_{\DDstarb})$, the number of $\DDstarb$ candidates in channel $i$,
to compute the average efficiency as a function of $m_{\DDstarb}$,

\begin{equation}
\epsilon^{\cal{B}}(m_{\DDstarb}) = \frac{\sum_{i=1}^{n} N_i(m_{\DDstarb})}{\sum_{i=1}^{n} \frac{N_i(m_{\DDstarb})}{\epsilon^{\cal{B}}_i(m_{\DDstarb})}},
\end{equation}
where $n$ is the number of decay modes. In this case we have eight \DstarzDzb chanels
(1-4 with $\Dstarz \to \Dz \gamma$ and $\Dstarz \to \Dz \piz$) and two  \DstarpDm channels (13, 14).
Representative values of $\epsilon^{\cal{B}}$, computed at a mass of 4.5 \gevcc, are displayed in 
Table~\ref{tab:pury}.
The sample sizes for the Cabibbo-suppressed decay modes (15, 16, and 17 in Table~\ref{tab:list_dstar}) 
are very small (32 events total)
and comprise 14\% of the  $\DstarpDm$sample. The efficiency for these decay channels has not been directly computed; 
instead, these  modes 
are treated as having the mean efficiency of the Cabibbo-allowed channels 13 and 14.
The ten \DDstarb channels, after correcting for efficiency and branching fractions,
have yields that are consistent within the statistical errors.

\begin{figure}[!htb]
\begin{center}
\includegraphics[width=10cm]{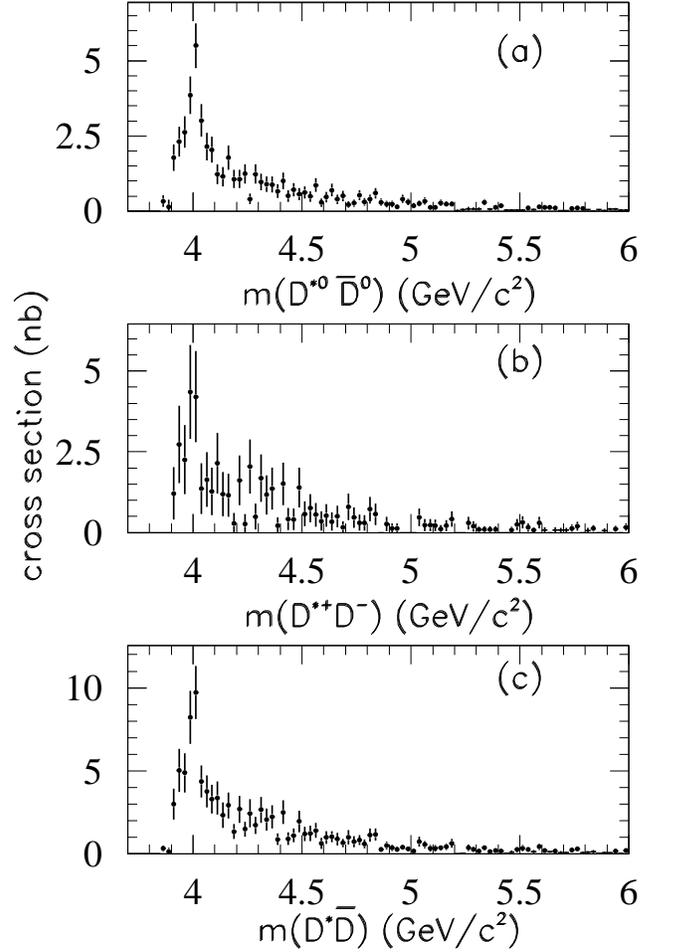}
\caption{Cross section for $e^+ e^- \to$ (a) \DstarzDzb, (b) \DstarpDm and (c) \DDstarb combined. The error bars correspond
to statistical errors only.} 
\label{fig:fig4}
\end{center}
\end{figure}

The $\DDstarb$ cross section is computed using
\begin{equation}
\sigma_{e^+e^-\to \DDstarb}(m_{\DDstarb}) = \frac{dN/dm_{\DDstarb}}{\epsilon^{\cal{B}}(m_{\DDstarb})
  d{\cal{L}}/dm_{\DDstarb}}, 
\end{equation}
where $dN/dm_{\DDstarb}$ is the background-subtracted yield.
The differential luminosity is computed as~\cite{benayoun}
\begin{equation}
\frac{d{\cal{L}}}{dm_{\DDstarb}} = {\cal{L}} \frac{2m_{\DDstarb}}{s} \frac{\alpha}{\pi
  x}(\ln(s/m_e^2)-1)(2-2x+x^2), 
\end{equation}
where $s$ is the square of the 
$e^+ e^-$ center-of-mass energy, $\alpha$ is the fine-structure constant, $x = 1 - m_{\DDstarb}^2/s$,
$m_e$ is the electron mass, and $\cal{L}$ is the
integrated luminosity of 384~\invfb.
The cross sections for \DstarzDzb, \DstarpDm, and combined \DstarzDzb and \DstarpDm are shown
in Fig.~\ref{fig:fig4}. A clear $\psi(4040)$ resonance is observed.

The systematic uncertainties on the cross sections, 10.9\% for \DstarzDzb and 9.3\% for \DstarpDm,
include uncertainties for particle identification, tracking, photon and \piz reconstruction efficiencies,
background estimates, branching fractions, and 
a potential inaccuracy in the
simulation of extraneous tracks, photon and \piz candidates.
The uncertainty due to the ISR selection has been estimated by narrowing
the $\MM$ allowed range to $0.7\ \gevcccc$. All contributions are added in quadrature. 
Systematic uncertainties are summarized in Table~\ref{tab:sys_dstard}. 
\begin{table}[tbp]
\caption{Systematic errors, given as fractional errors
expressed in \%, in the evaluation of the \DDstarb cross section.}
\label{tab:sys_dstard}
\begin{center}
\vskip -0.2cm
\begin{tabular}{lcc}
\hline \noalign{\vskip2pt}
Effect & $D^{*0} \bar D^0$ & $D^{*+} D^-$  \cr
\hline
Background subtraction &  2.6 & 3.0  \cr
Branching fractions & 7.4  & 4.6 \cr
$\MM$ cut & 2.2 & 0.0 \cr
Particle identification & 1.8 & 2.1 \cr
Tracking efficiency & 2.2 & 3.3 \cr
Extraneous tracks & 5.7 & 5.7 \cr
$\pi^0$ and $\gamma$ reconstruction efficiency& 3.4 & 3.0 \cr
Extraneous $\pi^0$ and $\gamma$ & 0.5 & 0.8 \cr
\hline
Total & 10.9 & 9.3 \cr
\hline
\end{tabular}
\end{center}
\end{table}

The \DstarzDzb and \DstarpDm cross sections have similar features and consistent yields. 
Integrating the cross sections from threshold to 6 \gevcc, 
we obtain
\begin{equation}
 \frac{\sigma(\DstarpDm)}{\sigma(\DstarzDzb)} = 0.95 \pm 0.09_{stat} \pm 0.10_{syst},
\end{equation}
\noindent
consistent with unity. In this calculation systematic errors related to the $\MM$ selection criteria and tracking efficiency 
have been ignored because they largely cancel in the ratio.
\section{Study of the {\boldmath \DstarDstarb} system}

\begin{figure}[!htb]
\begin{center}
\includegraphics[width=10cm]{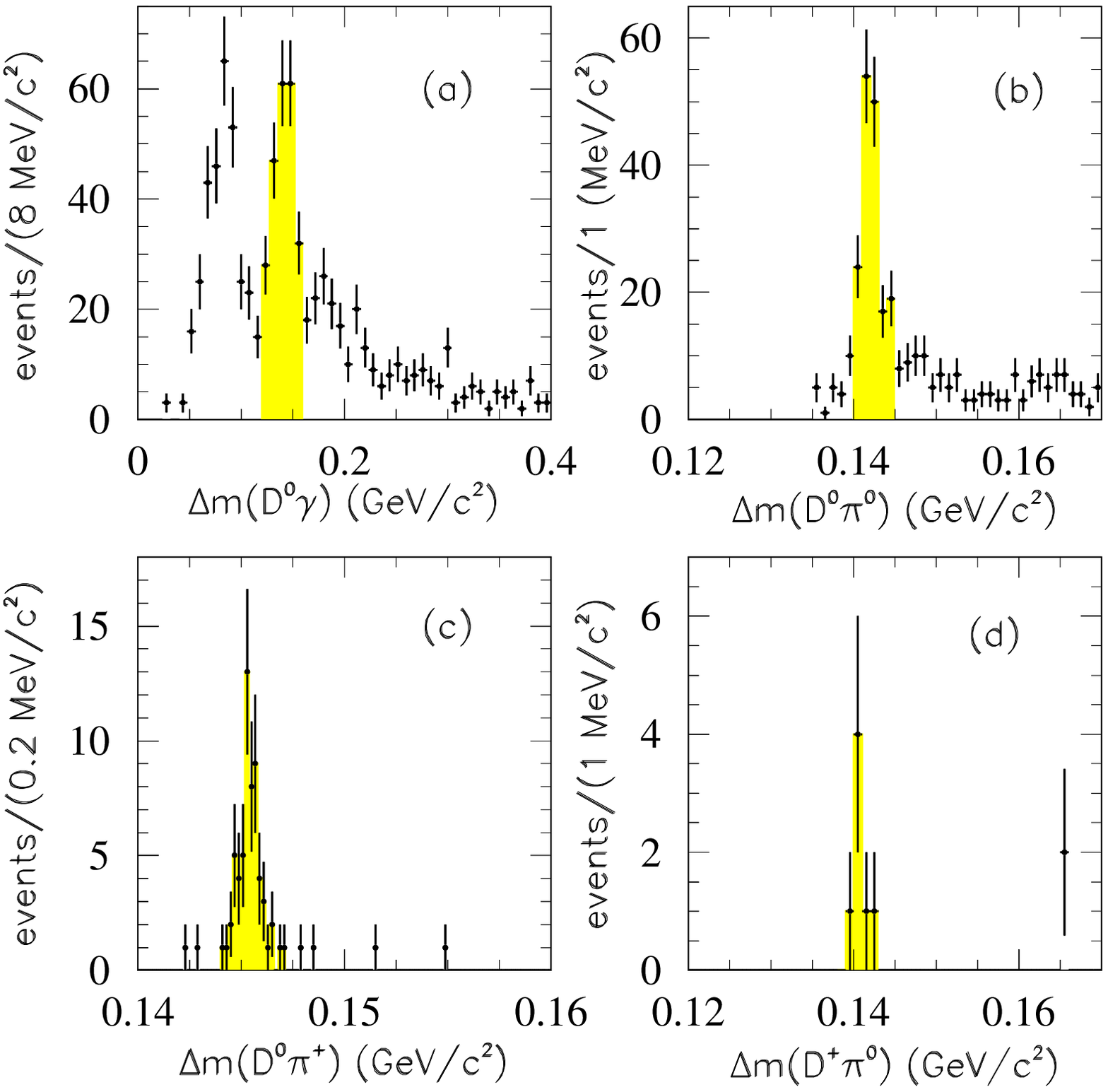}
\caption{$\Delta m$ distributions for \DstarDstarb candidates after applying the $| \MM| <1\ \gevcccc$ and
$m(\DstarDstarb)<6$ \gevcc selections,
for (a) $\Dstarz \to \Dz \gamma$,
(b) $\Dstarz \to \Dz \piz$, (c) $\Dstarp \to \Dz \pip$ with $\Dz \to \Km \pip$, and
(d) $\Dstarp \to \Dp \piz$ with $\Dp \to \Km \pip \pip$. The shaded regions indicate the ranges used to select
the $D^*$ signals.}
\label{fig:fig5}
\end{center}
\end{figure}

A similar analysis is carried out for $\DstarDstarb$ channels. 
Figure~\ref{fig:fig5} shows 
the $\Delta m$
distributions for $\DstarDstarb$ candidates 
with $|  \MM |  < 1\ \gevcccc$ and $\DstarDstarb$ masses below 6 \gevcc. 
The peak at threshold in Fig.~\ref{fig:fig5}(a) is due to background from $\Dstarz \to \Dz \piz$ where
one $\gamma$ from the low momentum $\piz$ is lost.

\begin{figure}[!htb]
\begin{center}
\includegraphics[width=10cm]{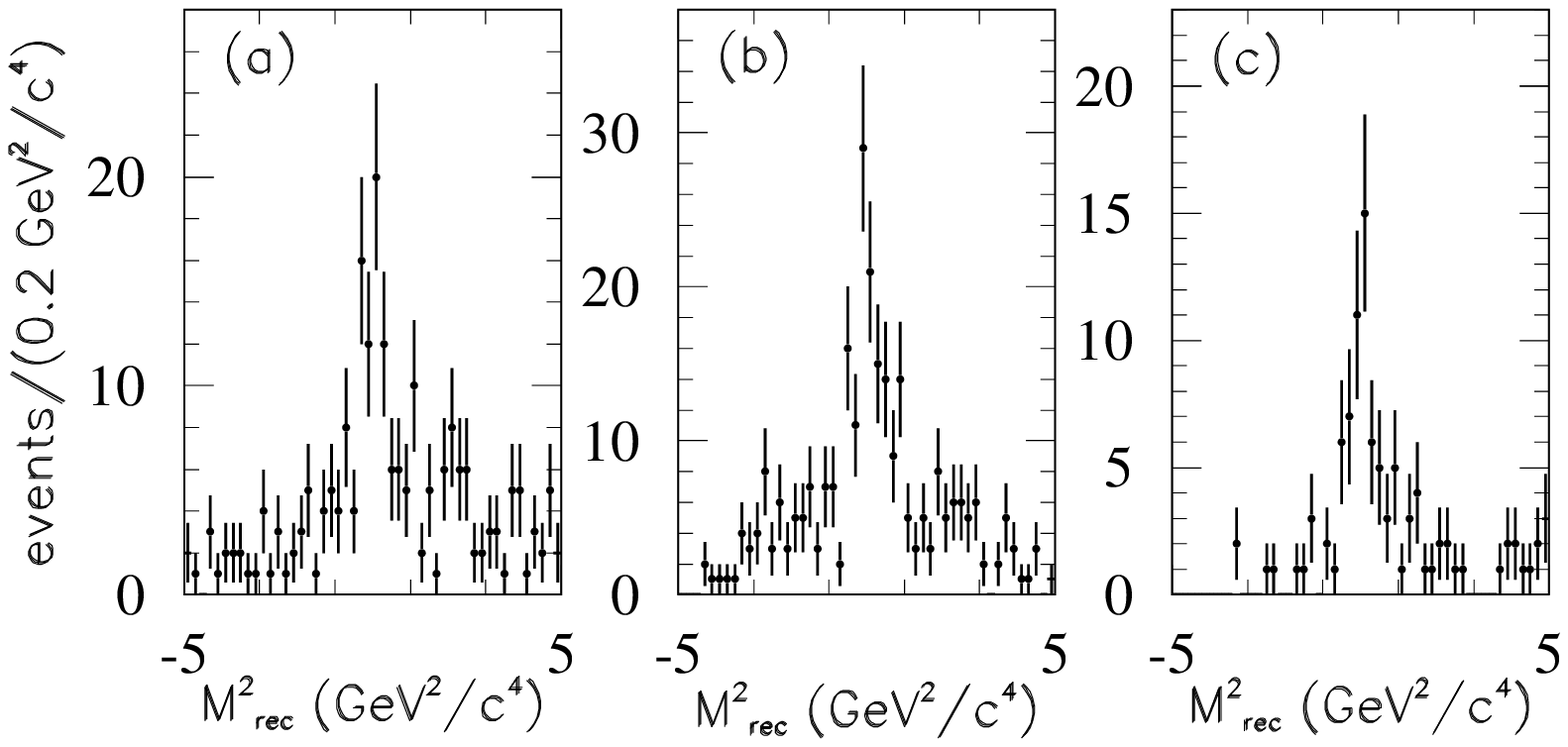}
\caption{$\MM$ distributions for \DstarzDstarzb for (a) $\Dstarz \to \Dz \gamma$, $\Dstarzb \to \Dzb \gamma$,
(b) $\Dstarz \to \Dz \piz$, $\Dstarzb \to \Dzb \gamma$, and  (c) $\Dstarz \to \Dz \piz$, $\Dstarzb \to \Dzb \piz$.}
\label{fig:fig6}
\end{center}
\end{figure}

We select the two \Dstar candidates and 
reject candidates reconstructed in any of the modes listed in the ``veto'' column in Table~\ref{tab:list_dstar}. 
Figure~\ref{fig:fig6} shows the \DstarzDstarzb $\MM$ distributions for channels 10-12.

\begin{figure}[!htb]
\begin{center}
\includegraphics[width=10cm]{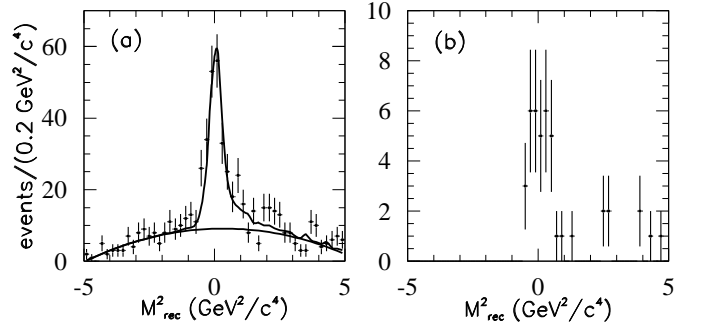}
\caption{$\MM$ distributions for (a) \DstarzDstarzb  and (b) \DstarpDstarm. The curve in (a) is the result from the fit 
described in the text.}
\label{fig:fig7}
\end{center}
\end{figure}

The total \DstarzDstarzb and  \DstarpDstarm $\MM$ distributions are shown in Fig.~\ref{fig:fig7}.
The number of background events for the \DstarzDstarzb channel is estimated by fitting the $\MM$ distribution. The fit is performed
using a $2^{nd}$-order polynomial for the background and a signal $\MM$ lineshape obtained from Monte Carlo simulations
that reflect the composition of the data. The number of ISR candidates and purities are also 
summarized in Table~\ref{tab:pury}.
The \DstarpDstarm final state has a background consistent with zero.

\begin{figure}[!htb]
\begin{center}
\includegraphics[width=8cm]{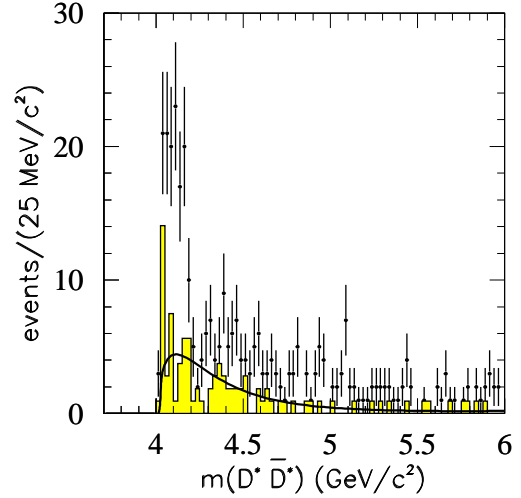}
\caption{\DstarDstarb mass spectrum. The shaded histogram is obtained
from the $\MM$ sidebands $-2.5 < \MM <-1.5$ and $1.5 < \MM < 2.5\ \gevcccc$. The curve is the 
result from the fit described in the text.} 
\label{fig:fig8}
\end{center}
\end{figure}

Due to the small \DstarpDstarm sample size, the charged and neutral mass spectra are summed in Fig.~\ref{fig:fig8}.
The \DstarDstarb mass spectrum shows unresolved peaks at $\psi(4040)$ and $\psi(4160)$ and an enhancement at the position of 
the $\psi (4400)$
~\cite{pdg}.  

The background is explored using events in the $\MM$ sideband regions $-2.5 < \MM <-1.5\ \gevcccc$ and 
$1.5 < \MM < 2.5\ \gevcccc$, and fitted using Eq.~(2). 
The \DstarDstarb mass spectrum for these events, normalized from the fit to the $\MM$ distribution,
is shown as the shaded histogram in Fig.~\ref{fig:fig8}.

\begin{figure}[!htb]
\begin{center}
\includegraphics[width=8cm]{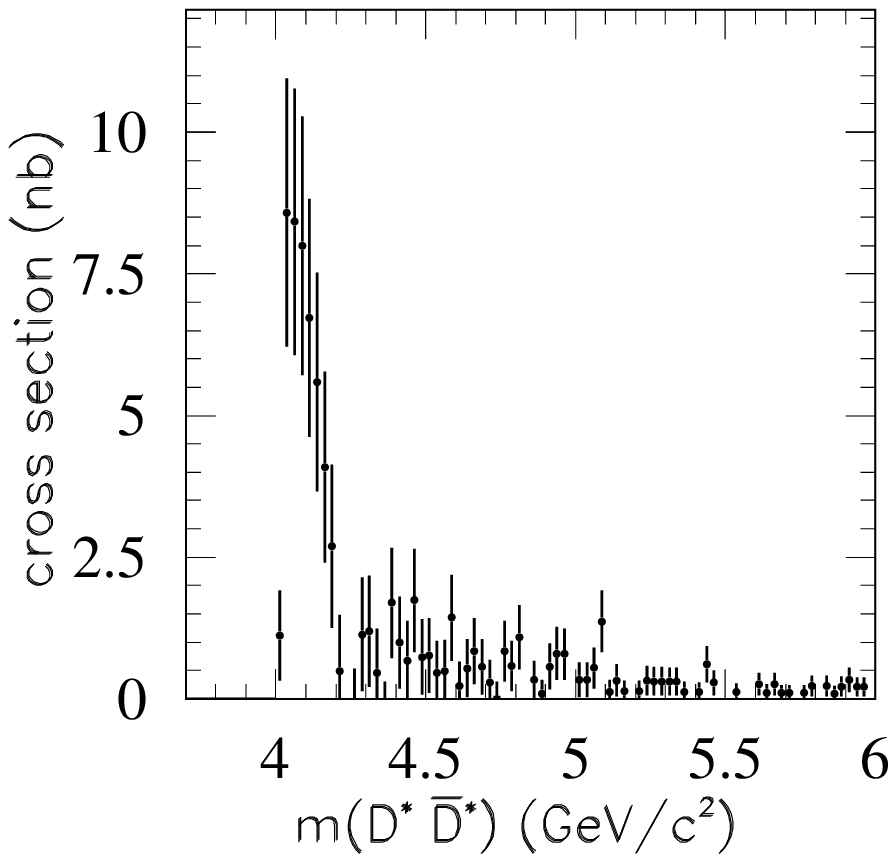}
\caption{Cross section $e^+ e^- \to \DstarDstarb$ for combined $\DstarzDstarzb$ and $\DstarpDstarm$. Error bars indicate the
statistical errors only.}
\label{fig:fig9}
\end{center}
\end{figure}

The \DstarDstarb cross section is calculated using the same method used to compute the 
\DDstarb cross section. 
The result, summed over the neutral and charged modes, is shown in Fig.~\ref{fig:fig9}.
All systematic uncertainties which have been taken into account for the  \DstarDstarb
mode are listed in Table~\ref{tab:sys_dstar2}; the overall uncertainty on 
the cross section is 12.4\%.
\begin{table}[tbp]
\caption{Systematic errors, given as fractional errors
expressed in \%, in the evaluation of the \DstarDstarb cross section.}
\label{tab:sys_dstar2}
\begin{center}
\vskip -0.2cm
\begin{tabular}{lc}
\hline
Effect & Fraction (\%) \cr
\hline
Background subtraction & 2.1 \cr
Branching fractions & 9.3 \cr
$\MM$ cut & 1.3 \cr
Particle identification & 2.8 \cr
Tracking efficiency & 2.6 \cr
Extraneous tracks & 5.7 \cr
$\pi^0$ and $\gamma$ reconstruction efficiency & 4.1 \cr
\hline
Total & 12.4 \cr
\hline
\end{tabular}
\end{center}
\end{table}

The \DstarDstarb cross section distribution exhibits a threshold
enhancement due to the superposition of the $\psi(4040)$ and $\psi(4160)$ resonances.

\section{The {\boldmath \DDb} mass spectrum}

\begin{figure}[!htb]
\begin{center}
\includegraphics[width=10cm]{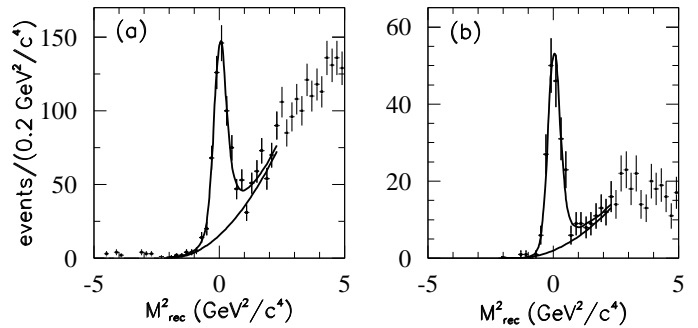}
\caption{$\MM$ distribution for (a) \DzDzb and (b) \DpDm. The curves are the results from the fits described in the text.}
\label{fig:fig10}
\end{center}
\end{figure}

\begin{figure}[!htb]
\begin{center}
\includegraphics[width=9cm]{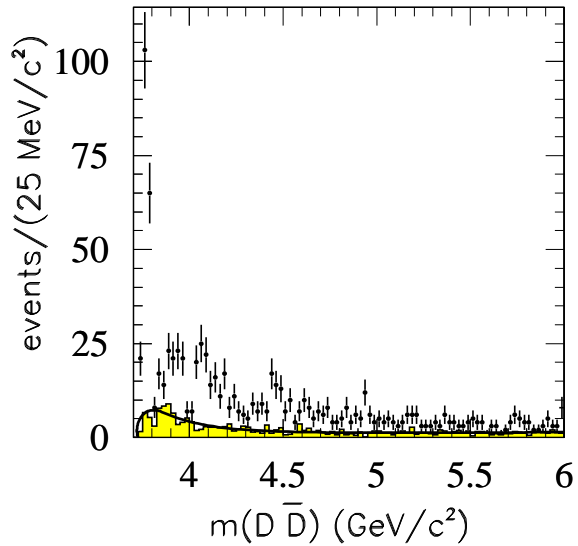}
\caption{\DDb mass distribution. The shaded histogram is obtained
from the $\MM$ sideband. The curve is the result from the fit described in the text.} 
\label{fig:fig11}
\end{center}
\end{figure}

In the selection of the \DzDzb sample we also apply the method of resolving ambiguous events having
an additional $\piz$ and/or $\gamma$. Here we veto all events that are ambiguous with channels
8-12, obtaining a rejection of 7.6~\% background events in the 
$| \MM |<1 \ \gevcccc$ region. 
No such procedure is applied to the \DpDm sample. The \DDb analysis is otherwise  
identical to that reported in Ref.~\cite{babar_dd}.
The resulting $\MM$ distributions for \DzDzb and \DpDm  channels are
shown in Fig.~\ref{fig:fig10}. The curves are the results from the fits performed using a $2^{nd}$ order polynomial for the background and a $\MM$ lineshape obtained from Monte Carlo simulations
that reflect the channel composition of the data. Again, the resulting event yields and purities are summarized in Table~\ref{tab:pury}. 
The combined \DDb mass spectrum  is shown in Fig.~\ref{fig:fig11}. The background is explored using events in the $\MM$ sideband regions $1.5 < \MM <3.5\ \gevcccc$ and fitted using Eq.~(2). 
This background, normalized from the fit to the $\MM$ distributions,
is shown as the shaded histogram in Fig.~\ref{fig:fig11}.
The features in the \DDb mass spectrum and the resulting \DDb cross section have been extensively discussed in our previous 
publication~\cite{babar_dd}.

\section{Fit to the Mass Spectra}

Unbinned maximum likelihood fits to the \DzDzb, \DpDm, \DstarzDzb, \DstarpDm, \DstarzDstarzb, and \DstarpDstarm mass spectra are performed.
We write the likelihood functions as
\begin{equation}
\begin{split}
L = f  \epsilon^{\cal{B}}(m) |  P(m) + c_1 W_1(m) e^{i \phi_1} + c_2 \sqrt{G(m)} e^{i \phi_2} + ... \\
+ c_n W_n(m)e^{i \phi_n} |  ^2 + B(m)(1 - f),
\end{split}
\end{equation}

\noindent
where $m$ is the $D^{(*)}\Dstarzzb$ mass, $c_i$ and
$\phi_i$ are free parameters, $W_i(m)$ are P-wave relativistic Breit-Wigner
distributions~\cite{pdg}, $P(m)$ represents the nonresonant contribution, $B(m)$ describes
the background, $\epsilon^{\cal{B}}(m)$ is the average efficiency, and $f$ is the signal fraction
fixed to the values obtained fitting the $\MM$ distributions. 

The parameters of the $\psi(4040)$,
$\psi(4160)$, and $\psi(4415)$ are fixed to the values reported in the Review of Particle Physics~\cite{pdg}. 
The parameters of the \psiprpr are fixed to the values obtained in our previous analysis of the \DDb system~\cite{babar_dd}.
The \DDb data require that we include the 3.9 GeV$/c^2$ structure,
as suggested in Ref.~\cite{eichten1},
which we parametrize empirically as the square root of
a Gaussian times a phase factor $\sqrt{G(m)} e^{i \phi_2}$. 
The parameters of the Gaussian are fixed to the values obtained in our previous analysis of the \DDb system: 
$m_{G(3900)} = 3943  \pm 17$ \mevcc, $\sigma_{G(3900)}=52 \pm 8$ \mevcc~\cite{babar_dd}. 
The shape of the nonresonant contribution $P(m)$ is unknown; we therefore parametrize it in a simple way
as
\begin{equation}
P(m) =C(m)(a + b m),
\end{equation}
where $C(m)$ is the phase space function for $D^{(*)}\Dstarzzb$, and $a$ and $b$ are free parameters.
Resolution effects have been ignored since the widths of the resonances are
much larger than the experimental resolution. 

Interference between the resonances and the
nonresonant contribution $P(m)$ is required to obtain a satisfactory description of the data. 
The size of the nonresonant production is determined by the fit.

\begin{figure}[!htb]
\begin{center}
\includegraphics[width=10cm]{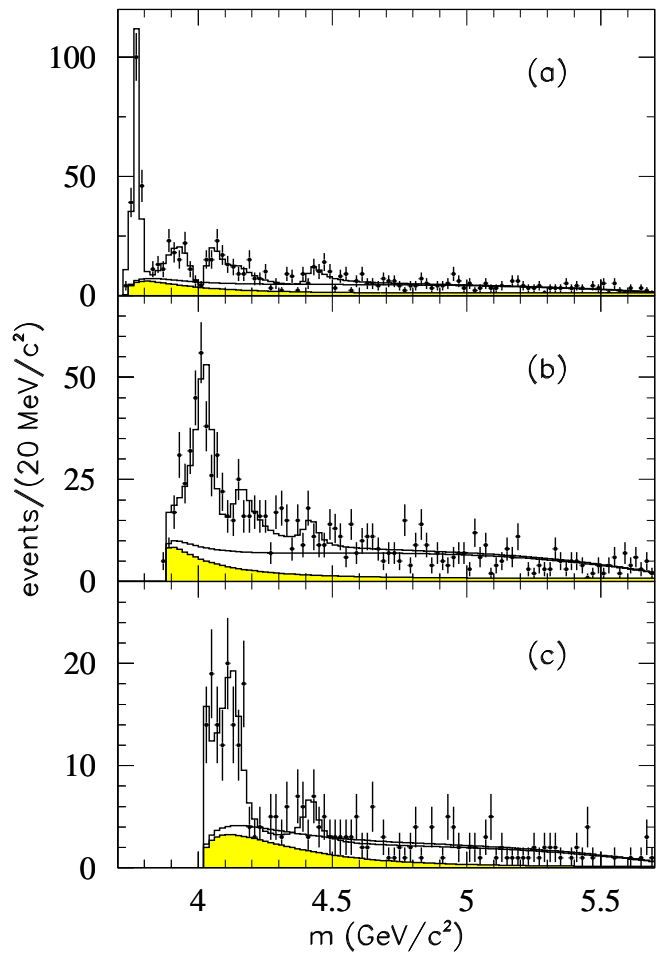}
\caption{Fits to the (a) \DDb, (b) \DDstarb, and (c) \DstarDstarb mass spectra. Data are represented with error bars, the curves represent the fitted functions. The shaded histogram corresponds to the smoothed incoherent background ($B(m)$) 
estimated from sidebands. The 
second smooth solid line represents the nonresonant contribution ($|  P(m) | ^2$).} 
\label{fig:fig12}
\end{center}
\end{figure}

The six \DzDzb, \DpDm, \DstarzDzb, \DstarpDm, \DstarzDstarzb, and \DstarpDstarm likelihood functions are computed
with different thresholds, efficiencies, purities, backgrounds, and numbers of contributing resonances appropriate for
each channel. The fits, summed over the charged and neutral final states, are shown in Fig.~\ref{fig:fig12};
they provide a good description of all the data. In the figure, the shaded areas indicate the background estimated by 
fitting the $\MM$ sidebands. The second smooth solid line represents the nonresonant contribution where we plot
$| P(m) |^2$, therefore ignoring the interference effects. 
The fraction for each resonant contribution $i$
is defined by the following expression:
\begin{equation}
f_i = \frac {|c_i|^2 \int |W_i(m)|^2 dm}
{\sum_{j,k} c_j c_k^* \int W_j(m) W_k^*(m) dm}.
\end{equation}
The fractions $f_i$ do not necessarily add up to 1 because of interference
between amplitudes. The error for each fraction has been evaluated by propagating the 
full covariance matrix obtained by the fit. The resulting fit fractions 
and phases are given in Table~\ref{tab:tabfrac}.
\begin{table*}[tbp]
\caption{ \DDb, \DDstarb, and \DstarDstarb fit fractions (in \%) and phases. Errors are statistical only.}
\label{tab:tabfrac}
\begin{center}
\vskip -0.2cm
\begin{tabular}{lcccccc}
\hline \noalign{\vskip2pt}
 & $\DDb$    & $\DDb$ & $\DDstarb$    & $\DDstarb$    & $\DstarDstarb$ & $\DstarDstarb$  \cr
Res. & fraction & phase  & fraction & phase  & fraction & phase \cr
\hline \noalign{\vskip2pt}
$|  P | ^2$ & 38.5 $\pm$ 7.1 &  0. & 49.9 $\pm$ 5.6 & 0. & 56.8 $\pm$ 9.2 & 0. \cr
$\psi(3770)$ & 31.3 $\pm$ 3.3 & $1.58$ $\pm$ 0.46 & & & &   \cr
$G(3900)$ & 23.9 $\pm$ 5.8 &  $5.46 $ $\pm$ 0.64 & & & &  \cr
$\psi(4040)$ & 31.2 $\pm$ 5.3 & $1.39$ $\pm$ 0.55  & 34.5 $\pm$ 6.0 &  $1.74$ $\pm$ 0.33 & 5.7 $\pm$ 4.4 &  $3.37 $ $\pm$ 0.48 \cr
$\psi(4160)$ & 3.1  $\pm$ 3.3 & $2.75$ $\pm$ 0.58 & 12.2 $\pm$ 3.8 &  $2.26$ $\pm$ 0.16 & 30.6 $\pm$ 7.3 &  $5.94 $ $\pm$ 0.33 \cr
$\psi(4400)$ & 2.0 $\pm$ 1.2 &  $3.38 $ $\pm$ 0.37 & 0.6 $\pm$ 0.7 &  $4.37$ $\pm$ 0.47 & 3.6 $\pm$ 2.4 &  $5.03 $ $\pm$ 0.45 \cr
\hline \noalign{\vskip2pt}
Sum & 130 $\pm$ 12 & & 97 $\pm$ 8 & & 97 $\pm$ 13  & \cr
\hline
\end{tabular}
\end{center}
\end{table*}

\section{Fit Fractions and Integrated Rates}

The fit gives corrected yields for each charmonium resonance. 
Since the fits have been performed independently for the neutral and charged modes, the weighted means of the fit
fractions are used.
These can be used to compute the integrated rates
for each resonance in the $\DDb$, $\DDstarb$, and $\DstarDstarb$ decay modes, which are 
reported in Table~\ref{tab:cross_res}.
\begin{table*}[tbp]
\caption{Integrated rates (in nb$\cdot$\mev) for $e^+ e^- \to \psi(4040)$, $e^+ e^- \to \psi(4160)$, and 
$e^+ e^- \to \psi(4400)$ in the $\DDb$, $\DDstarb$, and $\DstarDstarb$ decay modes. The first error is statistical, 
the second systematic.}
\label{tab:cross_res}
\begin{center}
\vskip -0.2cm
\begin{tabular}{lccc}
\hline
Decay mode &  $\psi(4040)$  &  $\psi(4160)$  &  $\psi(4400)$  \cr
\hline \noalign{\vskip2pt}
$\DDb$ & \ $11.0 \pm 1.8 \pm 5.6$  \  \  &  \  \ \  $1.0 \pm 1.3 \pm 1.0$ \ \   &  \  \ $0.5 \pm 0.3 \pm 0.1$   \cr
$\DDstarb$ & \  $46.6 \pm 7.0 \pm 4.9$  \ \  &  \  \ $13.8 \pm 4.4 \pm 1.5$ \  \  &  \  \ $0.6 \pm 0.8 \pm 0.1$ \cr
$\DstarDstarb$ &  \ $8.3 \pm 6.4 \pm 1.0$ \  \  &  \  \ $40.6 \pm 9.7 \pm 5.0$  \  \ &  \   $3.6 \pm 2.4 \pm 0.4$ \cr
\hline
\end{tabular}
\end{center} 
\end{table*}
The systematic errors take into account uncertainties on resonance parameters, Breit-Wigner lineshapes, branching fractions, and
background estimates. The nonresonant contribution has been parametrized in an alternative way, $P(m) =C(m)e^{a + b m}$.
Each resonance parameter has been varied according to its uncertainty, and the
meson radius used in the Blatt-Weisskopf damping factor~\cite{dump}, which is present 
in each relativistic Breit-Wigner term, has been varied between 0 and 2.5 ${\rm GeV}^{-1}$.
The amounts of backgrounds in the different final states have been varied according to their errors.
The 3.9 GeV$/c^2$ structure in the \DDb mass spectrum has been alternatively described by a P-wave 
relativistic Breit Wigner with free parameters. This effect dominates the systematic uncertainty on the $\psi(4040)$ rate
in the $\DDb$ mass spectrum. 
The deviations from the central value are added in quadrature.
Systematic effects also include the uncertainty on the total cross sections.

The corrected yields can also be used to compute the branching fraction ratios. The results are shown in Table~\ref{tab:br}
together with predictions of models: significant discrepancies are observed, expecially with the $^3P_0$ model~\cite{barnes}.

\begin{table*}[tbp]
\caption{Ratios of branching fractions for the three $\psi$ resonances. The first error is statistical, the second systematic.
Theoretical expectations are from the $^3P_0$ model~\cite{barnes}, $C^3$ model~\cite{eichten}, and $\rho K \rho$ model~\cite{swanson}.}
\label{tab:br}
\begin{center}
\vskip -0.2cm
\begin{tabular}{lccc}
\hline \noalign{\vskip2pt}
Ratio & \ \ measurement \ \ & \ \ \ $^3P_0$ \  & \ \ \ $C^3$ and $\rho K \rho$ \cr
\hline \noalign{\vskip2pt}
1) $\BR(\psi(4040)\to \DDb)/\BR(\psi(4040)\to \DDstarb)$ & 0.24 $\pm$ 0.05 $\pm$ 0.12 & 0.003 & 0.14~\cite{swanson} \cr 
2) $\BR(\psi(4040)\to \DstarDstarb)/\BR(\psi(4040)\to \DDstarb)$ & 0.18 $\pm$ 0.14 $\pm$ 0.03 & 1.0 &  0.29~\cite{swanson}\cr
3) $\BR(\psi(4160)\to \DDb)/\BR(\psi(4160) \to \DstarDstarb)$ & 0.02 $\pm$ 0.03  $\pm$ 0.02 & 0.46  & 0.08~\cite{eichten}\cr
4) $\BR(\psi(4160)\to \DDstarb)/\BR(\psi(4160) \to \DstarDstarb)$ &  0.34 $\pm$ 0.14 $\pm$ 0.05 & 0.011 & 0.16~\cite{eichten} \cr
5) $\BR(\psi(4400)\to \DDb)/\BR(\psi(4400) \to \DstarDstarb)$ & 0.14 $\pm$ 0.12 $\pm$ 0.03 & 0.025  & \cr
6) $\BR(\psi(4400)\to \DDstarb)/\BR(\psi(4400) \to \DstarDstarb)$ & 0.17 $\pm$ 0.25 $\pm$ 0.03 & 0.14 &\cr
\hline
\end{tabular}
\end{center} 
\end{table*}

\section{Limits on the decays ${\boldmath Y(4260) \to \DDstarb}$ and ${\boldmath Y(4260) \to \DstarDstarb}$}

The \DDstarb and \DstarDstarb mass spectra have been refit with an additional $Y(4260)$ resonance, which is allowed to
interfere with all the other terms.

The fit gives $Y(4260)$ fractions of ($2.2 \pm 2.9_{\rm stat} \pm 2.5_{\rm syst})\%$ and ($4.0 \pm 2.0_{\rm stat} \pm 4.2_{\rm syst})\%$ corresponding to 
$18 \pm 24_{\rm stat} \pm 21_{\rm syst}$ and $9 \pm 5_{\rm stat} \pm 10_{\rm syst}$ events for $Y(4260) \to \DDstarb$ and $Y(4260) \to \DstarDstarb$, respectively.
Systematic errors due to uncertainties on masses and widths of the $\psi(4040)$,
$\psi(4160)$, $\psi(4415)$, and $Y(4260)$ resonances are evaluated 
by varying the masses and widths by their uncertainty in the fit.
The amount of background in each final state is varied within its statistical error, and 
the meson radii in Breit-Wigner terms  are varied between 0 and 2.5 ${\rm GeV}^{-1}$.
Deviations from the central value are added in quadrature. 

These $Y(4260)$ yields in the \DDstarb and \DstarDstarb channels are used to 
compute the cross section times branching fraction, which can then be compared to our measurement from the 
$J/\psi\pi^+ \pi^-$ channel \cite{babar_Y}.
We obtain
\begin{equation}
\frac{\BR(Y(4260)\to \DDstarb)}{\BR(Y(4260)\to J/\psi \pi^+ \pi^-)} < 34,
\end{equation}
\noindent
and
\begin{equation}
\frac{\BR(Y(4260)\to \DstarDstarb)}{\BR(Y(4260)\to J/\psi \pi^+ \pi^-)} < 40,
\end{equation}
\noindent
at the 90\% confidence level.

\begin{figure}[!htb]
\begin{center}
\includegraphics[width=10cm]{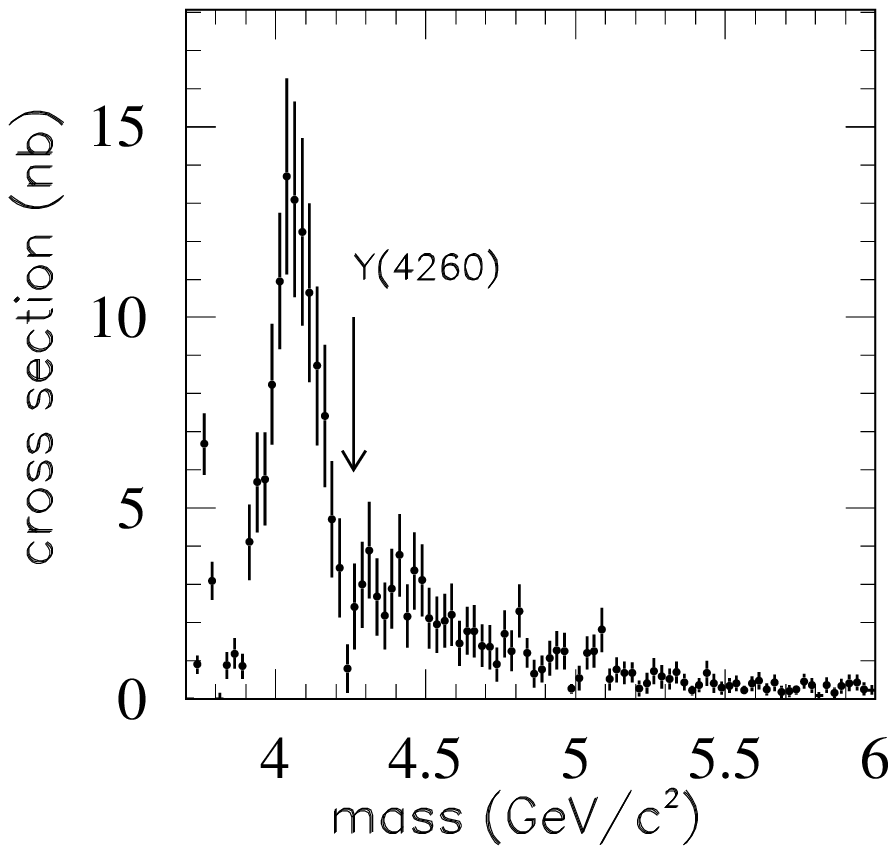}
\caption{Sum of $e^+ e^- \to  \DDb$, $e^+ e^- \to  \DDstarb$, and $e^+ e^- \to  \DstarDstarb$ cross sections.
The arrow indicates the position of the $Y(4260)$.} 
\label{fig:fig13}
\end{center}
\end{figure}

Using the $\DDb$ cross section measured in the earlier \babar \ work~\cite{babar_dd}, we obtain
the sum of the $e^+ e^- \to  \DDb$, $e^+ e^- \to  \DDstarb$, and $e^+ e^- \to  \DstarDstarb$ cross sections 
shown in Fig.~\ref{fig:fig13}: the arrow indicates the position of the $Y(4260)$, which falls in a 
local minimum, in agreement with the cross section measured for hadron production in $e^+e^-$ annihilation~\cite{pdg}. 

\section{Conclusions}

We have studied the exclusive ISR production of the  \DDb, \DDstarb, and \DstarDstarb systems.
The mass spectra show production of the $J^{PC}=1^{--}$ states \psiprpr,
$\psi(4040)$, $\psi(4160)$, and $\psi(4415)$. Fits to the mass spectra provide amplitudes and relative phases 
for the charmonium states, from 
which first measurements of branching fraction ratios are obtained. Finally, upper limits on  $Y(4260) \to \DDstarb$ and $Y(4260) \to \DstarDstarb$ decays are computed.

If the $Y(4260)$ is a $1^{--}$ charmonium state, it should decay predominantly to 
\DDb, \DDstarb, and \DstarDstarb~\cite{eichten,barnes}. Within the present limited data sample size, no evidence is found for $Y(4260)$ 
decays to $\DDb$,  \DDstarb, or \DstarDstarb.
Other explanations for the $Y(4260)$ have been proposed, such as a
hybrid, baryonium, molecule or tetraquark state. In the case of a hybrid state, the decay rates to \DDb, \DDstarb, and \DstarDstarb 
are expected to 
be small~\cite{Y4260-hybrid,iddir}. 

\section{Acknowledgements}
We are grateful for the 
extraordinary contributions of our \pep2\ colleagues in
achieving the excellent luminosity and machine conditions
that have made this work possible.
The success of this project also relies critically on the 
expertise and dedication of the computing organizations that 
support \babar.
The collaborating institutions wish to thank 
SLAC for its support and the kind hospitality extended to them. 
This work is supported by the
US Department of Energy
and National Science Foundation, the
Natural Sciences and Engineering Research Council (Canada),
the Commissariat \`a l'Energie Atomique and
Institut National de Physique Nucl\'eaire et de Physique des Particules
(France), the
Bundesministerium f\"ur Bildung und Forschung and
Deutsche Forschungsgemeinschaft
(Germany), the
Istituto Nazionale di Fisica Nucleare (Italy),
the Foundation for Fundamental Research on Matter (The Netherlands),
the Research Council of Norway, the
Ministry of Education and Science of the Russian Federation, 
Ministerio de Educaci\'on y Ciencia (Spain), and the
Science and Technology Facilities Council (United Kingdom).
Individuals have received support from 
the Marie-Curie IEF program (European Union) and
the A. P. Sloan Foundation.

\end{document}

%% file: authors_dec2008.tex
%
\author{B.~Aubert}
\author{Y.~Karyotakis}
\author{J.~P.~Lees}
\author{V.~Poireau}
\author{E.~Prencipe}
\author{X.~Prudent}
\author{V.~Tisserand}
\affiliation{Laboratoire d'Annecy-le-Vieux de Physique des Particules (LAPP), Universit\'e de Savoie, CNRS/IN2P3, F-74941 Annecy-Le-Vieux, France }
\author{J.~Garra~Tico}
\author{E.~Grauges}
\affiliation{Universitat de Barcelona, Facultat de Fisica, Departament ECM, E-08028 Barcelona, Spain }
\author{L.~Lopez$^{ab}$ }
\author{A.~Palano$^{ab}$ }
\author{M.~Pappagallo$^{ab}$ }
\affiliation{INFN Sezione di Bari$^{a}$; Dipartimento di Fisica, Universit\`a di Bari$^{b}$, I-70126 Bari, Italy }
% ...
\collaboration{The \babar\ Collaboration}
\noaffiliation

%% file: BABAR-PUB-08057.bbl
\begin{thebibliography}{99}
\bibitem{conj}
Charge conjugate states are implied throughout this work.
\end{thebibliography}
